\documentclass{aa}
\usepackage{epsfig}
\usepackage{times}
\newcommand{\etal}{{\it et al. }}
\newcommand{\hz}{high redshift }

\begin{document}
\thesaurus{11(11.01.2; 11.17.3); 13(13.25.2)}
\title{X-ray luminous radio-quiet high redshift QSOs in the ROSAT
All-Sky Survey}
\author{Xue-Bing Wu$^{1,2}$, Norbert Bade$^3$, Volker Beckmann$^3$}
 \institute{1. Beijing Astronomical Observatory, Chinese Academy of
Sciences, Beijing 100012, China\\
 2. CAS-PKU Joint Beijing Astrophysical Center, Beijing 100871, China\\
3. Hamburger Sternwarte, Gojenbergsweg 112, D-21029 Hamburg, Germany}
\date{\it Accepted on 03/19/1999 for publication in A\&A Main Journal }
\offprints{X.-B. Wu (wuxb@class1.bao.ac.cn)}
\maketitle

\markboth{X.-B. Wu \etal: X-ray luminous radio-quiet high redshift
QSOs in the RASS}{}

\begin{abstract}

X-ray luminous radio-quiet high redshift QSOs are
rare and can be used for the investigation of several important
astronomical questions. We have conducted a large area survey for
radio-quiet high redshift QSOs in the ROSAT All-Sky Survey, from 
which QSO
candidates are
selected using the
digitized objective prism spectra from the Hamburg Quasar Survey. The 22
candidates with Galactic
latitudes larger than $35^o$ were observed with the 2.16m telescope at
Xinglong station of Beijing Astronomical Observatory.
Among the 19 new QSOs in our sample, six are radio-quiet QSOs with
redshifts larger than 1.3 and three of them have redshifts
larger than 2. Thus we have doubled the number of known
X-ray luminous, radio-quiet high redshift QSOs in the surveyed sky area.
The distribution of $f_{1.4GHz}/f_{1keV}$ of radio-loud and
radio-quiet QSOs in this area shows two well
separated peaks. This can be taken as an evidence for different emission
mechanisms of the observed X-rays in the two subgroups.

\keywords{galaxies: active -- quasars: general--X-ray: galaxies}

\end{abstract}

\section{Introduction}

High redshift ($\rm z > 1.8$) QSOs, especially radio-quiet objects,  are
rare in the ROSAT All-Sky Survey (RASS, Voges \etal, 1996).
A comparison of the radio-loud and radio-quiet QSOs detected in the
RASS indicated that the X-ray detection rate of radio-quiet QSOs drops
much faster towards higher redshift than that of radio-loud QSOs
(Yuan \etal 1998). The catalogue of  Yuan \etal (1998) (see Table 1 in
that paper) contains 18 radio-quiet QSOs with $\rm z > 2$
detected in the RASS. In the northern sky with galactic
latitude $\rm b>35^o$, only 20 QSOs with $z>1.3$ and four QSOs with $z>2$
have been known in the literature so far.

At high redshifts the rest frame energy range is shifted to
higher energies, e.g., for $\rm z = 2.5$ the ROSAT/PSPC
(sensitive
between 0.1 and 2.4\,keV) receives photons emitted between 0.35 and
8.4\,keV. Only upper limits exist for radio-quiet QSOs above 100\,keV
whereas
several radio-loud objects with blazar properties have been detected
above 100 keV (e.g. von Montigny \etal 1995).
From several high redshift QSOs low energy cut-offs have been
found,
which are interpreted as 
due to the strong low energy absorption probably caused by
the material near the QSO itself (Elvis \etal 1998). Higher resolution
spectroscopy with future X-ray telescopes, e.g., AXAF, XMM,
 and 
Astro-E,
will be able to quantify the contribution of metals to this proposed strong
absorption.
Low energy cut-offs have mainly been discovered in radio-loud
objects, only some tentative
 evidence for low energy cutoffs are known in several
radio-quiet QSOs. However, the statistics
 above $\rm z >2.5$ is poor; only
three radio-quiet objects
are included in a previous study (Fiore \etal 1998).

In addition, high redshift QSOs found in the RASS belong to the
most luminous members of
 this class, making the
upper end of the luminosity function. It
is known from the magnification bias (Borgeest \etal 1991)
that such objects have high probability to be gravitationally
magnified. This is particularly true for radio-quiet objects, which seem to 
be the more common but intrinsically
X-ray faint population. Among the radio-quiet $\rm z > 2$ QSOs in the RASS,
two are known gravitationally lensed objects. 
HE1104--1805 ($\rm z = 2.4$) is a
double QSO detected in Hamburg (Wisotzki \etal 1993). The
other object is RX\,J0911.4+0551 ($\rm z = 2.8$, Bade \etal 1997),
a gravitationally lensed system with at least four QSO images.

The full sky coverage and the sensitivity of the RASS make it the 
most appropriate
for finding these rare objects. 
Radio-loud QSOs have
been selected mainly during the identifications of bright radio sources.
Projects which may be able to select
radio-quiet QSOs down to $\rm B = 18\,mag$ on large sky areas are under
way
(e.g. SDSS, see Kent 1994; Hamburg/ESO Survey, see Wisotzki \etal 1996);
but up to now results are only available for comparatively small and
patchy areas of the sky.
Therefore studies using correlations between existing AGN catalogues and 
the
RASS to determine the relation between radio bright and quiet QSOs in
the X-rays are probably affected by selection effects.

In order to enlarge the current sample of X-ray luminous radio-quiet
high redshift QSOs,
we start this survey project to select this kind of QSOs in the 
RASS
using optical observations. Our study will also verify whether some
selection
effects are serious in the above mentioned correlation
studies.

\section{High redshift QSO candidates selection}

We used the
digitized objective prism spectra from the Hamburg Quasar Survey
(HQS, Hagen \etal 1995) for the selection of QSOs with high redshifts.
Currently the Hamburg identification project of RASS sources (Bade
\etal 1998) covers about 10,200\,deg$^{2}$ of the northern high galactic
latitude sky ($\rm |b| > 20^o$). This survey uses photo plates which
are sensitive between 5400\,\AA\ and the
atmospheric limit at $\rm \sim 3400\AA$. Thus the 
Ly-$\alpha$ line
can be detected between $\rm 1.8 < z < 3.2$.
If the CIV\,1549
emission line is discernible in the objective prism spectrum, too, the
reliability of the classification is very high ($>90\%$). If only one
strong
emission line is visible, other strong emission lines can be source of
the line
(e.g.. CIII 1909, MgII 2798, [OIII]5007) and smaller redshifts are
possible. Follow-up observations of the QSO candidates are thus necessary.

The classification of AGN candidates in the Hamburg identification
project
of RASS sources is based mainly on their blue continuum which can be
well
described by a power law. For the selection of QSOs with higher
redshifts we
inspected the objective prism spectra by eye, looking for strong
emission lines. 
If the magnitude of the AGN candidate is near the plate limit, it is
difficult 
to distinguish emission lines from noise.
Another problem for number counts is the varying detection limit 
of the objective prism plates. Taking all effects into 
account, $\rm B = 18$ is a good approximation of the detection limit 
for strong emission lines.
Following these procedures we finally selected 22 QSO candidates
in the northern sky with galactic latitude $\rm b > 35^o$. 

In the 1st to 7th columns of Table 1, we list their ROSAT names,
coordinates of the
X-ray sources and their optical counterparts, offsets between the X-ray
and optical
positions, X-ray fluxes, B-band magnitudes and 1.4GHz radio fluxes
from NVSS (Condon \etal 1998).
The X-ray fluxes between 0.1 and 2.4\,keV are
calculated by assuming a power law with photon index $-1.8$
and only galactic neutral hydrogen absorption. 

The follow-up observations were split into two observing
campaigns in the spring of 1998 
when the area of 300 HQS plates covering  6500$\,\mbox{deg}^{2}$
was observable. 
Because varying\\
 weather conditions prevented us from observing the
high redshift QSO candidates uniformly, the surveyed area is only a rough
estimate.

\section{Spectroscopic observations of \hz QSO candidates}
We have performed the spectroscopic observations on the 22
selected \hz QSO candidates
using the 2.16m telescope at Xinglong station of Beijing
Astronomical Observatory.
The exposure time varied from 10 to 70 minutes for each object depending 
on weather conditions
and object brightness.
The spectra of these \hz QSO candidates were obtained after the
standard sky subtraction and relative
flux calibration using the MIDAS
software package developed at the European Southern Observatory (Banse \etal
1983).
The observation dates, exposure times, types and redshifts of these
\hz candidates are given in the 8th to 11th columns of Table 1. The
12th column gives the rest frame luminosity
 $\rm L_{X}$ between $\rm 0.1 - 2.4\,keV$.
For the
determination of
the luminosity
 $\rm H_{0} = 50\,km\,s^{-1}\,Mpc^{-1}$ and $\rm q_{0} = 0$
are assumed.

\tabcolsep 1.32mm
 \begin{table*}
 \caption[]{ Spectroscopic observation results of high redshift QSO
candidates
  \label{basistabel}}
\begin{tabular}{lccrrlrcrlcl}\\ \hline
\multicolumn{1}{c} {ROSAT names} & X-ray coordinates &Opt. coordinates&
Offset & $f_x^{*}$
&B & $f_{1.4G}$ & Date & Exp.  & Type & z & L$_x^{**}$ \\
& R.A. (2000.0)~~  Decl.&R.A. (2000.0)~~  Decl.&$('')$  &  & &(mJy) & &(s)&
&
 & \\
\hline
RX J0923.2+4602 &09 23 12.6 ~~46 02 42& 09 23 12.7 ~~46 02 42 & 2
& 4.9&18.4 & 11.7& 97/12/07 & 4200 & QSO & 0.729 & 1.87 \\
RX J0959.8+0049 & 09 59 48.9 ~~00 49 27 &09 59 46.9 ~~00 49 16 &33 &3.5
& 19 & &97/12/07 & 3600 & QSO & 2.243 & 26.9  \\
RX J0959.8+5942 & 09 59 49.3 ~~59 42 53& 09 59 48.5 ~~59 42 50 &8 &4.8&
17.5 & &97/12/07 & 1800 & Seyfert & 0.168 & 0.07  \\
RX J1059.8+0909 & 10 59 51.5 ~~09 09 15&  10 59 51.0 ~~09 09 05 &13& 4.0&
17.1 & &98/03/05  & 3000 & QSO & 1.683 & 13.6  \\
RX J1112.4+1101 & 11 12 25.5 ~~11 01 03& 11 12 25.4 ~~11 01 03 &2& 5.0&
18   & &98/05/24  & 900  & QSO & 0.636 & 1.37 \\
RX J1144.9+5434 &11 44 55.8 ~~54 34 58& 11 44 54.9 ~~54 34 51 &11 &12.7
& 18.4 & &98/03/05  & 1800 & QSO & 0.437 & 1.44  \\
RX J1208.3+5240 & 12 08 22.3 ~~52 40 40&  12 08 22.3 ~~52 40 12 &29
&13.0& 17.1 & 41.1& 98/03/05  & 1800 & QSO & 0.435 & 1.46  \\
RX J1243.8+0828 & 12 43 52..3 ~~08 28 25&  12 43 52.5 ~~08 28 26 &3 &
3.3& 17.6 & &98/05/24  & 900  & QSO & 0.384 & 0.28  \\
RX J1259.8+3423$^a$ &12 59 48.7 ~~34 23 25& 12 59 48.9 ~~34 23 19 &4 & 6.0&
16.9 & 11.8&98/05/24  & 600  & QSO & 1.376 & 11.7 \\
RX J1353.0+2947 & 13 53 00.0 ~~29 47 42& 13 52 59.1 ~~29 47 39 &14&
23.3& 18   & &98/05/24  & 1100 & QSO & 0.436 & 2.63 \\
RX J1404.1+0937 & 14 04 10.4 ~~09 37 45& 14 04 10.7 ~~09 37 45 &3
& 9.1& 17.3 & 20.1&98/05/24  & 600  & QSO & 0.437 & 1.03 \\
RX J1410.2+0811 & 14 10 13.6 ~~08 11 29 & 14 10 13.7 ~~08 11 28 &1 &5.2 &
19   & &98/05/24  & 800  & Seyfert & 0.088 & 0.02 \\
RX J1425.0+2749 &14 25 03.0 ~~27 49 21& 14 25 02.6 ~~27 49 10 &12 & 2.6&
18.8 & & 98/03/05  & 1800 & QSO & 2.346 & 22.8 \\
RX J1428.9+2710 & 14 28 54.0 ~~27 10 44& 14 28 53.0 ~~27 10 39 &17
& 4.7 & 17.8 & 7.2 &98/05/24  & 900  & QSO & 0.443 & 0.55 \\
RX J1441.2+3450 & 14 41 17.3 ~~34 50 53& 14 41 17.6 ~~34 50 52 &2 &
20.0& 16.8 & &98/05/24  & 800  & QSO & 0.352 & 1.34  \\
RX J1451.9+7214 &14 51 53.6 ~~72 14 41 & 14 51 54.1 ~~72 14 44 &6 &4.1 &
18   & &98/03/05  & 600  & QSO & 0.749 & 1.67 \\
RX J1514.3+4244 &15 14 20.4 ~~42 44 39& 15 14 20.4 ~~42 44 44 &5 &
15.9& 18.0 & 6.6 &98/05/24  & 600  & Seyfert & 0.152 & 0.18 \\
RX J1541.2+7126 & 15 41 16.3 ~~71 26 06 & 15 41 15.2 ~~71 25 58 &9 &3.9 &
18   & &98/03/05  & 900  & QSO & 1.418 & 8.26  \\
RX J1548.3+6949 & 15 48 18.7 ~~69 49 30& 15 48 16.7 ~~69 49 33 &11
& 8.3& 18.2 & 33.6& 98/05/24  & 900  & QSO & 0.375 & 0.66 \\
RX J1604.6+5714 &16 04 38.6 ~~57 14 37& 16 04 37.3 ~~57 14 36 &11
&7.3& 17.3 & 496.8& 98/05/24  & 800  & QSO & 0.725 & 2.75  \\
RX J1616.9+3621& 16 16 55.3 ~~36 21 33& 16 16 55.5 ~~36 21 34 &2
&0.7 & 16.9 & 330.9 &98/03/05  & 1200 & QSO & 2.259 & 5.50 \\
RX J1701.4+3511 &17 01 25.3 ~~35 11 50& 17 01 24.6 ~~35 11 56 &19 & 4.7 &
18   & &98/03/05  & 900  & QSO & 2.115 & 30.5  \\

\hline

\end{tabular}
\\ \vskip 0cm
\noindent Notes: $^{*)}$ $f_x$ is in unit of
$10^{-13}\,erg\,cm^{-2}\,s^{-1}$;
$^{**)}$ L$_x$ is the luminosity between $\rm 0.1 - 2.4\,keV$ in units
of
$\rm 10^{45}\,erg\,s^{-1}$;
$^{a)}$ A radio detected object but is formally radio quiet\\
\end{table*}

 \vskip -.2cm
\begin{figure*}
 \vskip -3.5cm
 \hskip -1.5cm
\rule{0.4pt}{4cm}
 \psfig{file=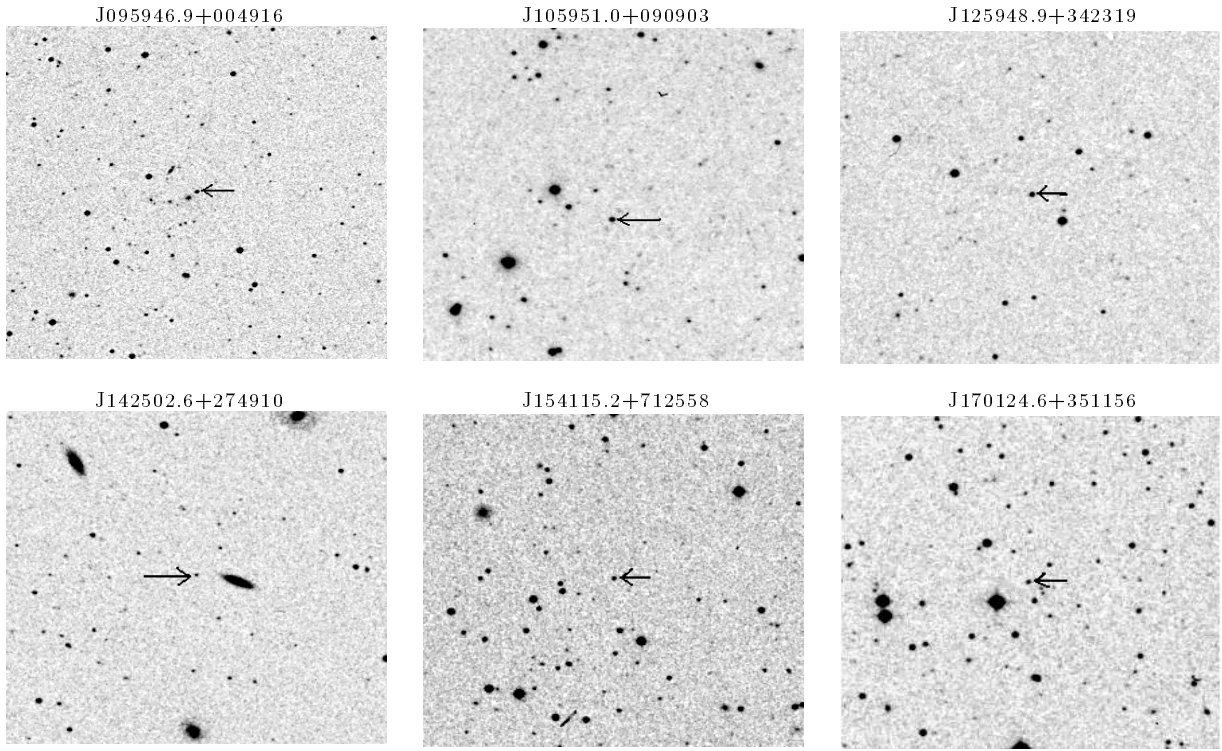,width=21.0cm,height=24.7cm,angle=0}
  \vskip -11cm
\caption{The $8'\times 8'$ POSS-I images of six new radio-quiet, X-ray
luminous quasars with redshift
larger than 1.3. The arrow points to the new quasar in each image.}

\end{figure*}

\begin{figure*}
\vskip -2.5cm
 \hskip -1.8cm
\rule{0.4pt}{4cm}
 \psfig{file=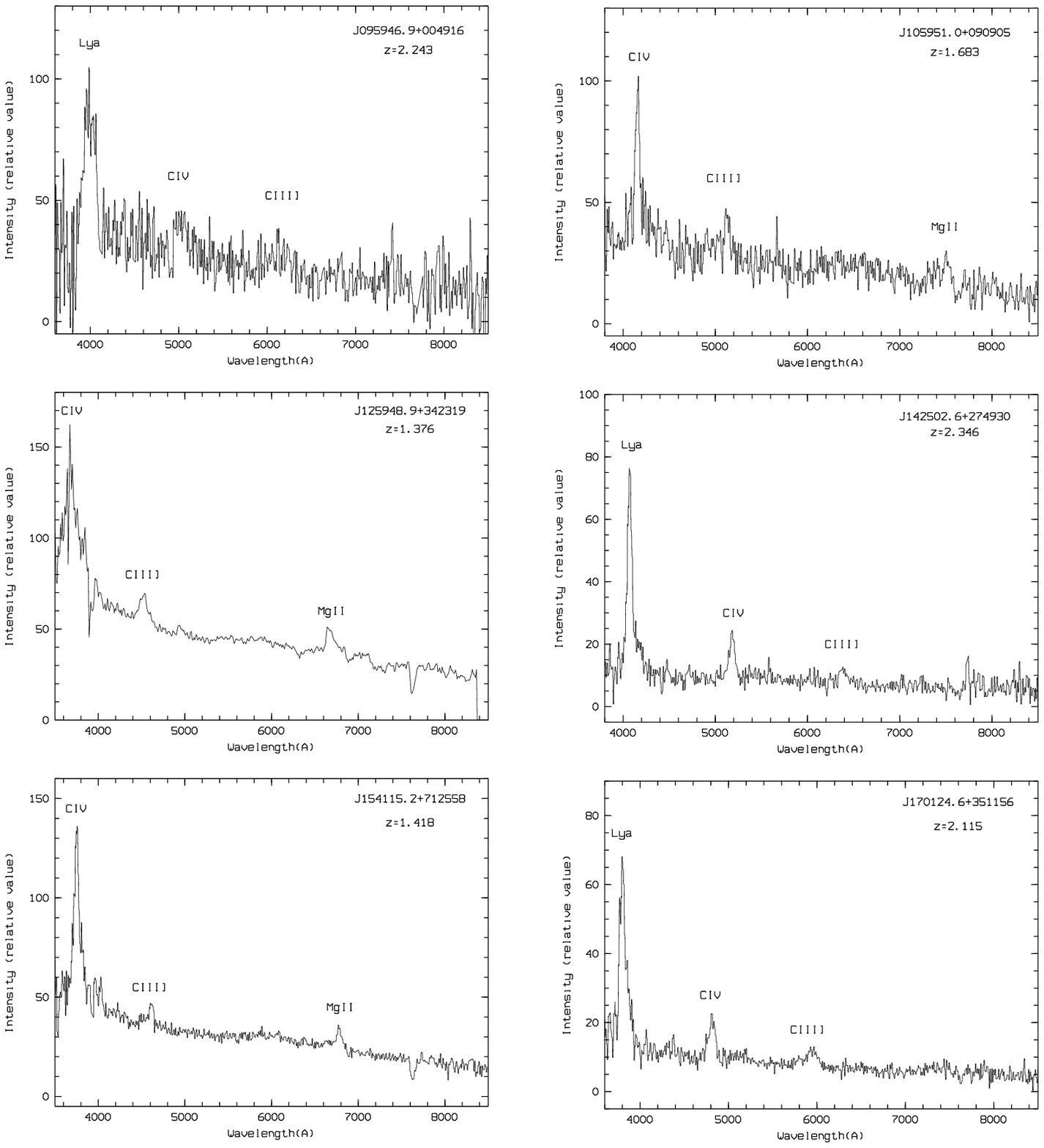,width=23cm,height=19.2cm,angle=0}
 \vskip -4.0cm
\caption{Spectra of 6 new high redshift radio-quiet, X-ray luminous
quasars. Three most significant emission lines in each spectrum are
labeled.}

\end{figure*}

Table 1 shows that
seven objects are QSOs with redshift larger
than 1, and four of them have redshift larger than 2. 
12 candidates are
QSOs with redshifts between 0.3 and 1.
Three candidates are low redshift Seyfert galaxies.
If the strongest emission line in the objective prism 
spectrum  is $\rm
Ly{\,\alpha}$,  redshift of the object is larger than 2.
If the strongest line is CIV or MgII in the prism spectrum, the redshift
is larger than 1 or between 0.3 and 1, respectively. However,
in several cases the emission
line in the prism spectrum is produced by a strong forbidden
emission line of Seyfert galaxies, such as [OII] or [NeV].
This is the reason why we also identified three Seyfert
galaxies.

We checked the radio fluxes of our \hz candidates with three 
main radio catalogues,
namely GB 5GHz catalogue (Gregory \etal 1996), NVSS 1.4GHz catalogue
(Condon \etal 1998)
and FIRST 1.4GHz catalogue (White \etal 1997). According to the
definition of radio-loud AGNs, namely $\rm R>10$
with $R$ being the ratio of 5GHz radio flux and the B-band optical flux
(Kellermann \etal 1989), we found that eight objects
in our sample are radio-loud (see Table 1). 
Among them six objects are weak radio sources and only two new quasars have
strong radio emission. RX J1616.9+3621 with $\rm z=2.259$, was found to
be positionally coincident with
a radio-loud object FIRST J161655.5+362134. 
Another new
QSO, RX J1604.6+5714 with $\rm z=0.725$, is coincident with 87GB
160335.3+572233.
One weak radio source (RX J1259.8+3423), which does not match the
definition  of radio-loud AGNs, 
and other 13 sources with no previous radio detection are most probably
radio-quiet objects.
Table 1 shows that five new QSOs have $\rm L_x (0.1 - 2.4\,keV)$\\
$\rm >10^{46}\,erg\,s^{-1}$,
and all of them are radio-quiet QSOs with $\rm z>1.3$. Another radio-quiet 
QSO with $\rm z>1.3$, RX J1541.2+7126, also has $\rm L_x 
=8.26 \times 10^{45}\,erg\,s^{-1}$. Therefore, these six new radio-quiet 
QSOs are really X-ray luminous objects.
In Figure 1 and 2, we give the finding charts and spectra of these six new
QSOs. 
The finding charts
were extracted from the Palomar (POSS-I) Digitized Sky-Survey .

Table 1 lists 18 sources having offsets less than $15''$
between
X-ray and optical positions.
They are all the most nearby optical sources to
the X-ray positions, so that we suggest that they are
the most likely true identifications of the X-ray sources. Four
candidates have X-ray to optical offsets larger than $15''$, which 
are not
uncommon for ROSAT detection at the 
detection limit. Here we
 discuss briefly each of these cases.
For {\it RX J0959.8+0049}, the optical counterpart we selected is $33''$
away from
the X-ray position and is identified as a quasar with $\rm z=2.243$.  Three
other optical sources (fainter than $O=19.5$) are found to be
closer to the X-ray
position, but all of them are red objects with $\rm O-E>2$. 
Although such colors are compatible with a M-type 
dwarf, the
optical
weakness makes this 
identification unplausible. Therefore,
 our identification
of this X-ray source with a $\rm z=2.243$ QSO is highly likely, though a 
future high resolution X-ray
image of this source is required to confirm this identification.
For {\it RX J1208.3+5240}, it is identified with a QSO with
$z=0.345$, which is $29''$ away
from the X-ray position. Another nearby optical source is fainter
than our candidate and is $38''$ far away from the X-ray source. Thus we
regard our identification is most likely correct.
For {\it RX J1428.9+2710}, its X-ray position is $17''$ away from 
 a QSO
with $z=0.443$. 
We note that another optical
source, is only $8''$ away from the
X-ray position. This source, with $O=19.2$ and $O-E=1.1$, is fainter
than our candidate QSO and is also point-like. Again future
high quality X-ray observations are needed to confirm our identification. 
Finally for {\it RX J1701.4+3511}, the optical source we selected is $19''$
away
from the X-ray position and was found to be a new QSO
with $z=2.115$. Other four nearby optical sources, with angular distances
from
$25''$ to $39''$ to the X-ray position, are faint and red objects with
$O-E>2$.
Therefore, we believe that our identification is reliable.

\section{Discussion}

 \begin{figure}
\vskip -0.2cm
\hspace*{0.4cm}
 \psfig{file= 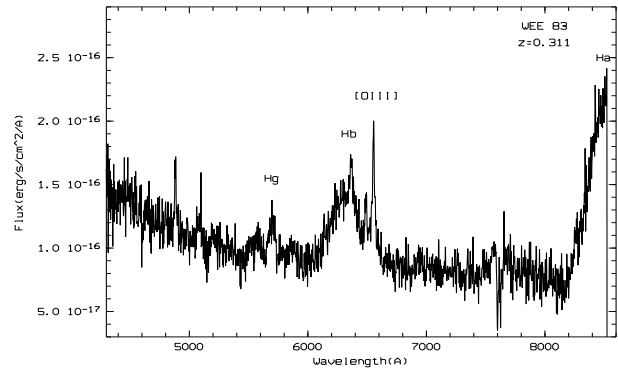,width=5.2cm,height=8.5cm,angle=270}
\caption{The slit spectrum of WEE 83. The emission lines of
$H_{\gamma}$, $H_{\beta}$ and [OIII] are evident and the
left wing of $H_{\alpha}$ is clearly shown,}
\end{figure}

\begin{figure*}
\vskip -1.5cm
\hspace*{-0.5cm}
\psfig{file= 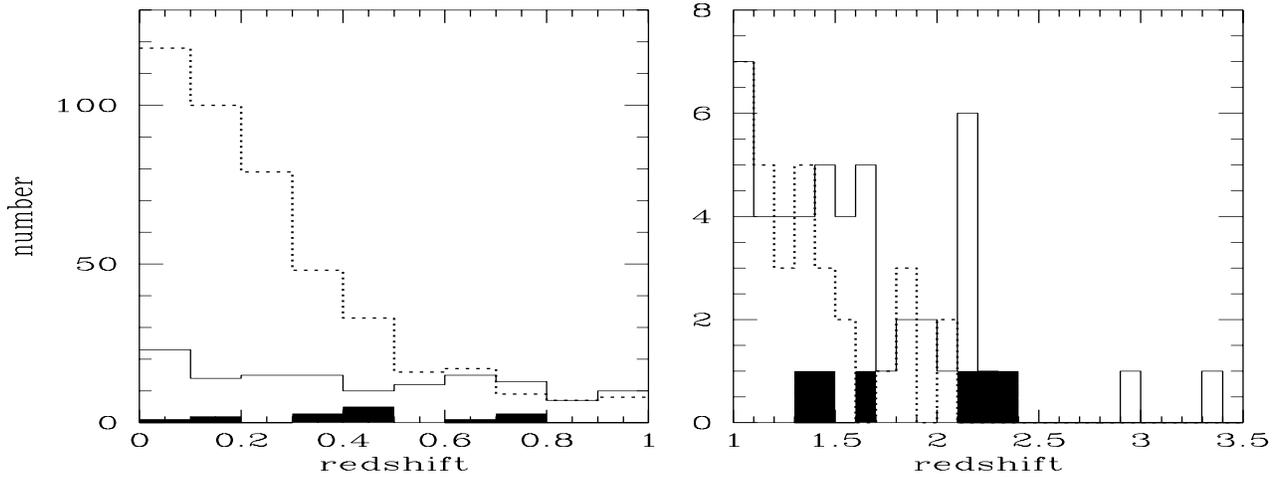,width=18.5cm,height=10cm,angle=0}
\vskip -1.9cm
\caption{The histogram of redshift distributions of previously known
QSOs detected by the RASS in the sky area with
$\rm b > 35^o$, and that of the newly discovered  AGNs in our sample. The
redshift
bin is 0.1. The solid and dotted
lines correspond to the previously known AGNs with 5GHz radio fluxes
larger than 20mJy and those with 5GHz radio
fluxes less than 20mJy, respectively. The shaded areas present
the
 newly discovered radio-quiet AGNs in our sample.}
\end{figure*}

Among the new radio-quiet QSOs discovered by us, six have redshifts
larger than 1.3 and three of them have
redshifts larger than 2. These identifications substantially
increase the number of such objects detected in the RASS.
In the sky area of high galactic latitudes  ($\rm
b > 35^o$), we have made a correlation between the catalogue
of identifications of RASS sources from the HQS Schmidt plates
and the catalogue of QSOs and AGNs (V\'eron-Cetty \& V\'eron 1996).
This area contains only 18 previously known
QSOs with 5GHz radio fluxes less than 20mJy
and redshifts larger than 1.3, and only four of them have redshifts larger
than 2. These four QSOs are RX J09190+3502, KUV 12461+2710, PG
1247+268 and WEE  83.  The QSO identified with RX J09190+3502 
was
discovered at Hamburg (Bade \etal
1995), but a confident identification was not made previously because an AGN
with low redshift was detected closer to the X-ray position. WEE 83  was
selected by Weedman
(1985) on objective prism plates, because it showed a strong emission line,
which was regarded as $\rm Ly{\,\alpha}$. In Figure 3 
we show the slit
spectrum of WEE 83 which we took in February 1998. Clearly we can see at 
least four
emission lines and the
average redshift of this object is 0.311 rather than 2.04.
Therefore, actually only two X-ray luminous and
radio-quiet QSOs with $\rm z>2$ in the sky
area of $\rm b > 35^o$ were known previously. In Figure 4 we show the
redshift distributions
of the previously known radio-loud and radio-quiet AGNs in the area with
$\rm b > 35^o$, and the redshift
distribution of the new radio-quiet AGNs in our sample. It is clear that
the radio-quiet QSOs are more numerous than
the radio-loud QSOs at lower redshift ($\rm z<0.7$);
most of the QSOs with higher redshift have high radio fluxes, and
radio-quiet QSOs with higher redshifts are relatively rare.
The new QSOs discovered by us represent a substantial contribution to
the number of radio-quiet, X-ray luminous QSOs with $\rm z>1.6$. 

A histogram of the ratio between the 1.4 GHz radio flux density
and the 1 keV X-ray flux density for the high redshift QSOs in our survey
area is given in
Figure 5. It includes 22 radio-loud QSOs and 9 radio-quiet QSOs with
$\rm z>1.6$ 
(an upper limit of 2.5mJy was adopted
for the sources with no radio detection in the NVSS catalogue).
It demonstrates again that the radio-loud objects with high redshift are
more numerous than the radio-quiet objects in the RASS,
however the amount of radio-quiet high redshift QSOs was obviously
underestimated in earlier studies.
This histogram shows clearly that the radio-loud QSOs and radio-quiet
QSOs are well separated and there
is a strong distinction between these two populations
of X-ray loud QSOs; it is clear from Figure 5 that all radio-loud high
redshift QSOs have $f_{1.4GHz}/f_{1keV}$
larger than $10^6$ while all radio-quiet QSOs have
$f_{1.4GHz}/f_{1keV}$ less than $10^6$. They seem to support the suggestion
that radio-loud QSOs and radio-quiet QSOs may have
different physical properties.
This has already been shown and discussed in earlier papers.
For example, the differences of the X-ray color and the X-ray spectra
of radio-quiet QSOs and radio-loud QSOs at high redshift have been found
by Bechtold \etal (1994), and these differences may imply that the
central engine
and the environment of these two kinds of QSOs are probably different.
It has also been found that the X-ray spectra of the radio loud high
redshift QSOs are significantly
harder compared to the radio quiet population (Schartel \etal 1996;
Brinkmann,
Yuan \& Siebert 1997).
It is thus possible that an additional spectral component is necessary in
order to explain the observations.
Such a component may be contributed by the beamed radiation 
from a highly relativistic plasma
jet 
aligned nearly to our direction (Wilkes \& Elvis 1987; Kollgaard 1994).

\begin{figure}
\vskip -0.3cm
\hspace{-0.3cm}
\psfig{file= 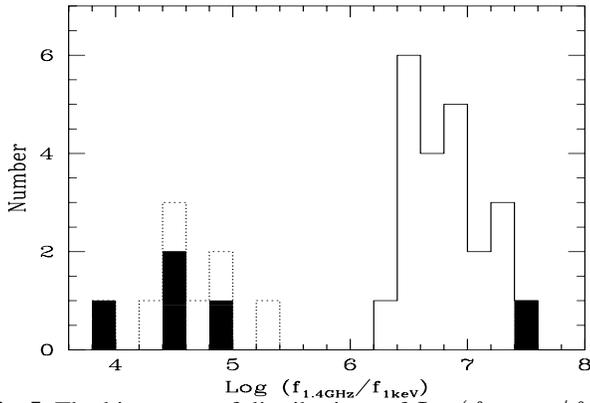,width=9cm,height=6cm,angle=0}
\vskip -0.5cm
\caption{The histogram of distributions of $Log(f_{1.4GHz}/f_{1keV})$
for QSOs with $\rm z > 1.6$ detected by the RASS in the sky area with $\rm b
> 35^o$. The solid and
dotted lines refer to the radio-loud QSOs and radio-quiet QSOs respectively.
The shaded area shows the contribution of the new QSOs in our sample. }
\end{figure}

For high redshift QSOs, the spectral region of the soft X-ray
excess
found with ROSAT is shifted
partially out of the observable window and 
with the spectral resolution of ROSAT it is difficult to
distinguish 
the soft excess and the power law component. However,  a strong
 soft X-ray excess 
with an extension to higher energies is  probably responsible for
the high 
X-ray fluxes of the
radio-quiet high redshift QSOs.
The origin of the soft X-ray excess is extensively
discussed in the literature, but no firm conclusion can be drawn currently
(Piro, Matt \&
Ricci 1997). Many authors have proposed
the inner regions of the accretion disk as emission region (e.g. Arnaud
\etal 1985; Saxton \etal 1993; Brunner \etal 1997).
Spectroscopic ASCA observations have shown that
 narrow-line 
Seyfert 1's and low redshift radio-quiet QSOs have generally steeper 
intrinsic hard
 X-ray continua (Brandt,
Mathur \& Elvis 1997; Reeves \etal 1997).
But whether the X-ray luminous, radio-quiet high redshift QSOs
have similar X-ray spectral properties as  narrow line 
Seyfert 1's (e.g. Boller, Brandt \& Fink 1996)
and low redshift radio-quiet QSOs is still unclear (see Vignali
 \etal 1998).
We expect that the future broad
 band spectroscopic observations
 on these 
QSOs using
 XMM and AXAF will
 lead to significant progress in understanding 
the X-ray emission mechanism
of them.

\begin{acknowledgements}
We thank Weimin Yuan for reading the manuscript critically and
the  
referee, Wolfgang Brinkmann, for helpful
comments. X.-B. Wu acknowledges the
support from the Director Funds of Beijing Astronomical Observatory.
This work is
based on the observations made by the 2.16m telescope at Xinglong
station of Beijing Astronomical Observatory.
\end{acknowledgements}

\end{document}